\documentclass[aps,pra,notitlepage,%
superscriptaddress,twocolumn]{revtex4-1}

\usepackage{natbib}
\usepackage{graphicx}
\usepackage{amssymb}
\usepackage{amsmath}
\usepackage{ifthen}
\usepackage{braket}
\usepackage{xcolor}
\usepackage{bm}
\usepackage{bbm}
\usepackage{comment}
\usepackage{siunitx}
\usepackage{float}
\usepackage{dcolumn}
\newcolumntype{d}[1]{D{.}{.}{#1}}
\usepackage{subfigure}

\usepackage{fancyvrb}

\definecolor{garrosgreen}{rgb}{0.1, 0.4, 0.1}
\definecolor{dartmouthgreen}{rgb}{0.05, 0.5, 0.06}
\definecolor{duelferred}{rgb}{0.7, 0.2, 0.1}
\definecolor{cambridgeblue}{rgb}{0.1, 0.3, 1.0}
\definecolor{oxfordblue}{rgb}{0.05, 0.2, 0.7}

\newcommand{\calO}{\mathcal{O}}

\newcommand{\dd}{\mathrm{d}}
\newcommand{\ii}{\mathrm{i}}
\newcommand{\ee}{\mathrm{e}}

\newcommand{\rmR}{\mathrm{R}}

\renewcommand{\Im}{\mathrm{Im}\,}

\usepackage{xcolor}
\usepackage{bbm}

\definecolor{light}{gray}{0.90}
\definecolor{darker}{gray}{0.50}
\definecolor{dark}{gray}{0.30}

\begin{document}

\title{Irreducible Three--Loop Vacuum--Polarization 
Correction in Muonic Bound Systems}

\author{Gregory S. Adkins}
\affiliation{Department of Physics and Astronomy, Franklin \& Marshall College,
Lancaster, Pennsylvania, 17604, USA}

\author{Ulrich D. Jentschura}
\affiliation{Department of Physics and LAMOR, Missouri University of Science and
Technology, Rolla, Missouri 65409, USA}

\begin{abstract}
Three-loop electronic vacuum-polarization corrections
due to irreducible diagrams
are evaluated for two-body muonic ions with nuclear
charge numbers $1 \leq Z \leq 6$.
The corrections are of order 
$\alpha^3 (Z\alpha)^2 \, m_r$, where $\alpha$ is the 
fine-structure constant and $m_r$ is the 
reduced mass. Numerically, the energy corrections
are found to be of the same order-of-magnitude as 
the largest of the order $\alpha^2 (Z \alpha)^6 m_r$
corrections, and are thus phenomenologically interesting.  
Our method of calculation eliminates 
numerical uncertainty encountered in other approaches.
\end{abstract}

\maketitle


%
%
\section{Introduction}

Energy corrections to bound states of 
muonic ions due to electronic 
vacuum-polarization effects are 
known to be numerically large,
due to the smallness of the 
generalized Bohr radius of the muonic ions.
For the $n$-loop energy shift $E^{(n)}$, one 
obtains the estimates
$E^{(n)} \sim \alpha^n (Z \alpha)^2 \, m_r $,
where $\alpha$ is the fine-structure constant, $Z$ is the nuclear charge  
number, and $m_r$ is the
reduced mass of the two-body system.

In particular, the two-loop vacuum-polarization 
contributions to the $2P$--$2S$ energy shift
in muonic hydrogen has recently 
been re-evaluated in Ref.~\cite{LaJe2024}.
From the irreducible two-loop diagrams,
one obtains a contribution of $1.25298$\,meV 
[see Eq.~(71) of Ref.~\cite{LaJe2024}],
while, from the reducible diagram,
one obtains $0.25495$\,meV [see Eq.~(72) of Ref.~\cite{LaJe2024}.
These results confirm the two-loop corrections 
obtained in Ref.~\cite{Pa1996mu}.
The sum of the two-loop contributions is 
about five times larger than the 
energy shift corresponding to the 
so-called proton radius puzzle~\cite{PaEtAl2010,Je2011aop1,Je2011aop2}.

Hence, it is of interest to
consider the three-loop vacuum polarization effect
for the bound states in 
muonic ions.
The scalar vacuum polarization function $\Pi^{(3)}(q^2)$ 
describing the irreducible three-loop
vacuum-polarization diagrams has
been discussed in its asymptotic 
(short-distance, high $q^2$) limit
in the context of the Gell--Mann Low
$\psi$ function (see Ref.~\cite{BaJo1969})
and for the Callan--Symanzik 
renormalization group (see Ref.~\cite{dRRo1974}).
The evaluation of three-loop vacuum-polarization
corrections in bound systems relies 
on knowledge of $\Pi^{(3)}(q^2)$ beyond the 
asymptotic (short-distance) regime.
The three-loop vacuum polarization spectral density
function ${\rm Im} \big [ \,\Pi^{(3)}_{\rm R}(q^2) \big ]$
has recently been evaluated analytically \cite{On2022}.
Using the spectral density, the full three-loop vacuum polarization
function $\Pi^{(3)}_{\rm R}(q^2)$ can be conveniently obtained 
using a subtracted dispersion  relation.

For muonic bound states, 
the evaluation of three-loop energy corrections has been
discussed in Refs.~\cite{KiNo1999prl,KiNo1999prd,%
IvKoKa2009}. Here, we are concerned with the 
evaluation of the irreducible three-loop
contribution, which, from the point of 
view of quantum field theory, probably constitutes the 
conceptually most interesting, gauge-invariant, 
subset of three-loop corrections.

This paper is organized as follows.
Some relevant properties
of the spectral function of three-loop vacuum-polarization
function are discussed in Sec.~\ref{sec2},
before we consider the numerical 
evaluation of the three-loop energy
shifts in Sec.~\ref{sec3}.
Conclusions are reserved for Sec.~\ref{sec4}.
We use natural units in this paper with 
$\hbar = c = \epsilon_0 = 1$.  In addition, we use the ``West-Coast'' convention
for the space-time metric: $g_{\mu\nu} = {\rm diag}(1,-1,-1,-1)$,
where $\mu,\nu = 0,1,2,3$.

\begin{figure*}[t!]
\begin{center}
\begin{minipage}{0.99\linewidth}
\begin{center}
\includegraphics[width=0.98\linewidth]{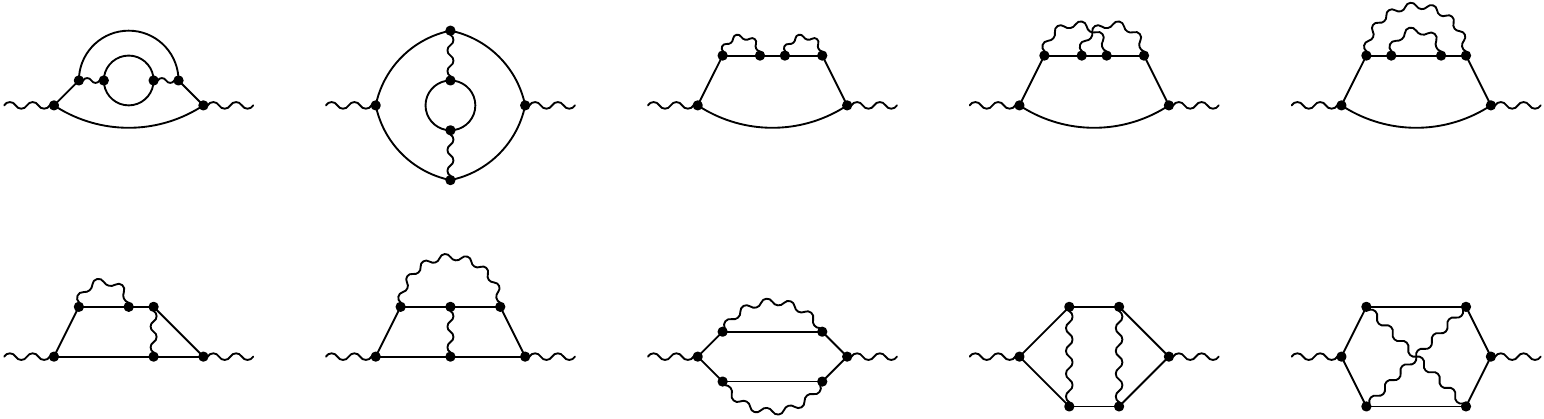}
\caption{\label{fig1}
The irreducible three-loop vacuum-polarization diagrams
are shown, with the fermion lines denoting the virtual 
electrons and virtual positrons.}
\end{center}
\end{minipage}
\end{center}
\end{figure*}

\begin{figure}[t!]
\begin{center}
\includegraphics[width=0.98\linewidth]{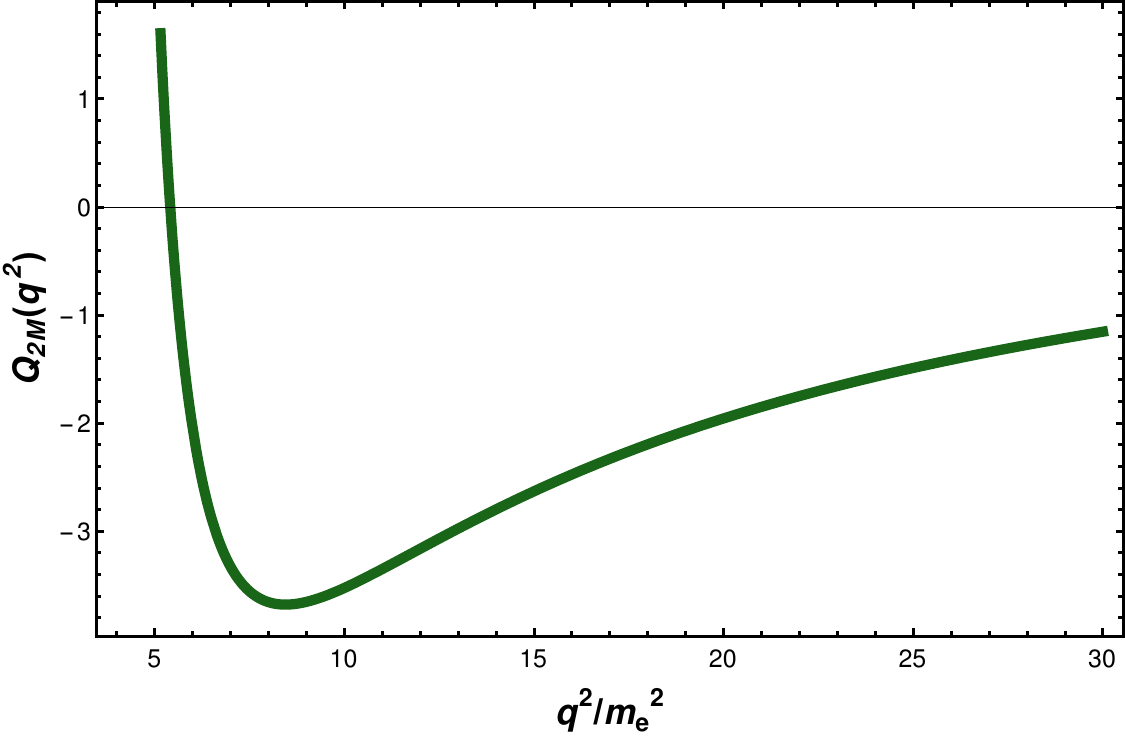}
\caption{\label{fig2}
The two-fermion threshold contribution 
to the imaginary part of the three-loop vacuum-polarization
function is given as $Q_{2m}(q^2)$ in Eq.~\eqref{ImP3R}.
It has an integrable singularity at the 
threshold $q^2 = (2 m_e)^2$ [see Eq.~\eqref{P2Mthr}].}
\end{center}
\end{figure}

\begin{figure}[t!]
\begin{center}
\includegraphics[width=0.98\linewidth]{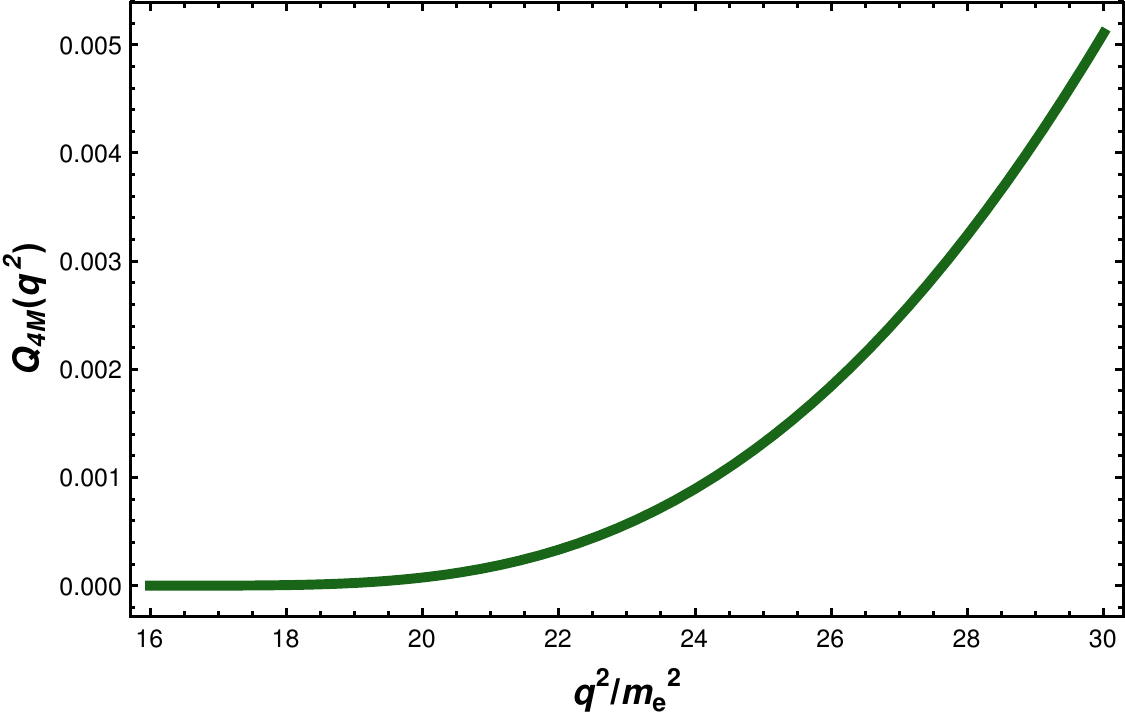}
\caption{\label{fig3}
Same as Fig.~\ref{fig2}, but for the
four-fermion threshold contribution $Q_{4m}(q^2)$.
It ramps up smoothly from the threshold at
$q^2 = (4 m_e)^2$ [see Eq.~\eqref{P4Mthr}].}
\end{center}
\end{figure}

\begingroup
\squeezetable
\def\arraystretch{1.1}
\begin{table*}
\begin{center}
\begin{minipage}{0.9\linewidth}
\begin{center}
\caption{\label{table1} Three-loop energy shifts due 
to irreducible three-loop diagrams are given
for muonic bound systems in the range $1 \leq Z \leq 6$
of nuclear charge numbers,
given in units of meV. 
The numerical uncertainties are $\pm 1$ in the least significant digit shown.}
\begin{tabular}{c@{\hspace{5ex}}c@{\hspace{5ex}}l@{\hspace{5ex}}%
l@{\hspace{5ex}}l@{\hspace{5ex}}l}
\hline
\hline
 \multicolumn{1}{c}{Bound System} & $n$
 & \multicolumn{1}{c}{$nS$} & \multicolumn{1}{c}{$nP$}
 & \multicolumn{1}{c}{$nD$} & \multicolumn{1}{c}{$nF$} \\
\hline
\multicolumn{1}{l}{$\mu$H} &
$n = 1$ & $ -2.5148 \times 10^{-2} $ & \multicolumn{1}{c}{---} & \multicolumn{1}{c}{---} & \multicolumn{1}{c}{---} \\
& $n = 2$ & $ -2.6243 \times 10^{-3} $ & $ -6.5673 \times 10^{-4} $ & \multicolumn{1}{c}{---} & \multicolumn{1}{c}{---} \\
& $n = 3$ & $ -7.5797 \times 10^{-4} $ & $ -2.0310 \times 10^{-4} $ & $ -1.0841 \times 10^{-5} $ & \multicolumn{1}{c}{---} \\
& $n = 4$ & $ -3.1704 \times 10^{-4} $ & $ -8.6759 \times 10^{-5} $ & $ -6.0104 \times 10^{-6} $ & $ -1.1495 \times 10^{-7} $ \\
\hline
\multicolumn{1}{l}{$\mu$D} &
$n = 1$ & $ -2.7476 \times 10^{-2} $ & \multicolumn{1}{c}{---} & \multicolumn{1}{c}{---} & \multicolumn{1}{c}{---} \\
& $n = 2$ & $ -2.8541 \times 10^{-3} $ & $ -7.7875 \times 10^{-4} $ & \multicolumn{1}{c}{---} & \multicolumn{1}{c}{---} \\
& $n = 3$ & $ -8.2386 \times 10^{-4} $ & $ -2.3863 \times 10^{-4} $ & $ -1.4109 \times 10^{-5} $ & \multicolumn{1}{c}{---} \\
& $n = 4$ & $ -3.4454 \times 10^{-4} $ & $ -1.0167 \times 10^{-4} $ & $ -7.7703 \times 10^{-6} $ & $ -1.6478 \times 10^{-7} $ \\
\hline
\multicolumn{1}{l}{$\mu^3$He} &
$n = 1$ & $ -1.6343 \times 10^{-1} $ & \multicolumn{1}{c}{---} & \multicolumn{1}{c}{---} & \multicolumn{1}{c}{---} \\
& $n = 2$ & $ -1.8071 \times 10^{-2} $ & $ -1.2009 \times 10^{-2} $ & \multicolumn{1}{c}{---} & \multicolumn{1}{c}{---} \\
& $n = 3$ & $ -5.0874 \times 10^{-3} $ & $ -3.1380 \times 10^{-3} $ & $ -7.0445 \times 10^{-4} $ & \multicolumn{1}{c}{---} \\
& $n = 4$ & $ -2.1163 \times 10^{-3} $ & $ -1.2868 \times 10^{-3} $ & $ -3.3621 \times 10^{-4} $ & $ -2.8701 \times 10^{-5} $ \\
\hline
\multicolumn{1}{l}{$\mu^4$He} &
$n = 1$ & $ -1.6550 \times 10^{-1} $ & \multicolumn{1}{c}{---} & \multicolumn{1}{c}{---} & \multicolumn{1}{c}{---} \\
& $n = 2$ & $ -1.8348 \times 10^{-2} $ & $ -1.2280 \times 10^{-2} $ & \multicolumn{1}{c}{---} & \multicolumn{1}{c}{---} \\
& $n = 3$ & $ -5.1618 \times 10^{-3} $ & $ -3.2020 \times 10^{-3} $ & $ -7.3004 \times 10^{-4} $ & \multicolumn{1}{c}{---} \\
& $n = 4$ & $ -2.1470 \times 10^{-3} $ & $ -1.3124 \times 10^{-3} $ & $ -3.4757 \times 10^{-4} $ & $ -3.0182 \times 10^{-5} $ \\
\hline
\multicolumn{1}{l}{$\mu^{6}$Li} &
$n = 1$ & $ -4.3559 \times 10^{-1} $ & \multicolumn{1}{c}{---} & \multicolumn{1}{c}{---} & \multicolumn{1}{c}{---} \\
& $n = 2$ & $ -5.6686 \times 10^{-2} $ & $ -4.7059 \times 10^{-2} $ & \multicolumn{1}{c}{---} & \multicolumn{1}{c}{---} \\
& $n = 3$ & $ -1.5350 \times 10^{-2} $ & $ -1.1362 \times 10^{-2} $ & $ -4.9511 \times 10^{-3} $ & \multicolumn{1}{c}{---} \\
& $n = 4$ & $ -6.3292 \times 10^{-3} $ & $ -4.5635 \times 10^{-3} $ & $ -2.0785 \times 10^{-3} $ & $ -3.8702 \times 10^{-4} $ \\
\hline
\multicolumn{1}{l}{$\mu^{7}$Li} &
$n = 1$ & $ -4.3712 \times 10^{-1} $ & \multicolumn{1}{c}{---} & \multicolumn{1}{c}{---} & \multicolumn{1}{c}{---} \\
& $n = 2$ & $ -5.6952 \times 10^{-2} $ & $ -4.7315 \times 10^{-2} $ & \multicolumn{1}{c}{---} & \multicolumn{1}{c}{---} \\
& $n = 3$ & $ -1.5418 \times 10^{-2} $ & $ -1.1421 \times 10^{-2} $ & $ -4.9946 \times 10^{-3} $ & \multicolumn{1}{c}{---} \\
& $n = 4$ & $ -6.3569 \times 10^{-3} $ & $ -4.5864 \times 10^{-3} $ & $ -2.0950 \times 10^{-3} $ & $ -3.9191 \times 10^{-4} $ \\
\hline
\multicolumn{1}{l}{$\mu^{9}$Be} &
$n = 1$ & $ -8.5559 \times 10^{-1} $ & \multicolumn{1}{c}{---} & \multicolumn{1}{c}{---} & \multicolumn{1}{c}{---} \\
& $n = 2$ & $ -1.2644 \times 10^{-1} $ & $ -1.1033 \times 10^{-1} $ & \multicolumn{1}{c}{---} & \multicolumn{1}{c}{---} \\
& $n = 3$ & $ -3.3769 \times 10^{-2} $ & $ -2.6421 \times 10^{-2} $ & $ -1.6351 \times 10^{-2} $ & \multicolumn{1}{c}{---} \\
& $n = 4$ & $ -1.3764 \times 10^{-2} $ & $ -1.0410 \times 10^{-2} $ & $ -6.2674 \times 10^{-3} $ & $ -1.9036 \times 10^{-3} $ \\
\hline
\multicolumn{1}{l}{$\mu^{10}$Be} &
$n = 1$ & $ -8.5699 \times 10^{-1} $ & \multicolumn{1}{c}{---} & \multicolumn{1}{c}{---} & \multicolumn{1}{c}{---} \\
& $n = 2$ & $ -1.2672 \times 10^{-1} $ & $ -1.1057 \times 10^{-1} $ & \multicolumn{1}{c}{---} & \multicolumn{1}{c}{---} \\
& $n = 3$ & $ -3.3842 \times 10^{-2} $ & $ -2.6482 \times 10^{-2} $ & $ -1.6410 \times 10^{-2} $ & \multicolumn{1}{c}{---} \\
& $n = 4$ & $ -1.3793 \times 10^{-2} $ & $ -1.0433 \times 10^{-2} $ & $ -6.2875 \times 10^{-3} $ & $ -1.9134 \times 10^{-3} $ \\
\hline
\multicolumn{1}{l}{$\mu^{10}$B} &
$n = 1$ & $ -1.4392 \times 10^{0} $ & \multicolumn{1}{c}{---} & \multicolumn{1}{c}{---} & \multicolumn{1}{c}{---} \\
& $n = 2$ & $ -2.3119 \times 10^{-1} $ & $ -2.0220 \times 10^{-1} $ & \multicolumn{1}{c}{---} & \multicolumn{1}{c}{---} \\
& $n = 3$ & $ -6.2396 \times 10^{-2} $ & $ -4.9895 \times 10^{-2} $ & $ -3.7513 \times 10^{-2} $ & \multicolumn{1}{c}{---} \\
& $n = 4$ & $ -2.5110 \times 10^{-2} $ & $ -1.9292 \times 10^{-2} $ & $ -1.3586 \times 10^{-2} $ & $ -5.7294 \times 10^{-3} $ \\
\hline
\multicolumn{1}{l}{$\mu^{11}$B} &
$n = 1$ & $ -1.4411 \times 10^{0} $ & \multicolumn{1}{c}{---} & \multicolumn{1}{c}{---} & \multicolumn{1}{c}{---} \\
& $n = 2$ & $ -2.3158 \times 10^{-1} $ & $ -2.0253 \times 10^{-1} $ & \multicolumn{1}{c}{---} & \multicolumn{1}{c}{---} \\
& $n = 3$ & $ -6.2507 \times 10^{-2} $ & $ -4.9987 \times 10^{-2} $ & $ -3.7610 \times 10^{-2} $ & \multicolumn{1}{c}{---} \\
& $n = 4$ & $ -2.5153 \times 10^{-2} $ & $ -1.9326 \times 10^{-2} $ & $ -1.3619 \times 10^{-2} $ & $ -5.7509 \times 10^{-3} $ \\
\hline
\multicolumn{1}{l}{$\mu^{12}$C} &
$n = 1$ & $ -2.2111 \times 10^{0} $ & \multicolumn{1}{c}{---} & \multicolumn{1}{c}{---} & \multicolumn{1}{c}{---} \\
& $n = 2$ & $ -3.7447 \times 10^{-1} $ & $ -3.2349 \times 10^{-1} $ & \multicolumn{1}{c}{---} & \multicolumn{1}{c}{---} \\
& $n = 3$ & $ -1.0354 \times 10^{-1} $ & $ -8.3630 \times 10^{-2} $ & $ -7.0401 \times 10^{-2} $ & \multicolumn{1}{c}{---} \\
& $n = 4$ & $ -4.1215 \times 10^{-2} $ & $ -3.1862 \times 10^{-2} $ & $ -2.4732 \times 10^{-2} $ & $ -1.3145 \times 10^{-2} $ \\
\hline
\multicolumn{1}{l}{$\mu^{13}$C} &
$n = 1$ & $ -2.2132 \times 10^{0} $ & \multicolumn{1}{c}{---} & \multicolumn{1}{c}{---} & \multicolumn{1}{c}{---} \\
& $n = 2$ & $ -3.7490 \times 10^{-1} $ & $ -3.2384 \times 10^{-1} $ & \multicolumn{1}{c}{---} & \multicolumn{1}{c}{---} \\
& $n = 3$ & $ -1.0368 \times 10^{-1} $ & $ -8.3739 \times 10^{-2} $ & $ -7.0519 \times 10^{-2} $ & \multicolumn{1}{c}{---} \\
& $n = 4$ & $ -4.1266 \times 10^{-2} $ & $ -3.1902 \times 10^{-2} $ & $ -2.4772 \times 10^{-2} $ & $ -1.3177 \times 10^{-2} $ \\
\hline
\hline
\end{tabular}
\end{center}
\end{minipage}
\end{center}
\end{table*}
\endgroup

%
%
\section{Analytic Derivation}
\label{sec2}

Our calculations make extensive use of the spectral function 
of three-loop vacuum polarization,  which was recently obtained 
expressed in terms of generalized polylogarithms in Ref.~\cite{On2022}.
The irreducible three-loop diagrams are shown in Fig.~\ref{fig1}.
We use the vacuum-polarization
function in the conventions of Ref.~\cite{LaJe2024} and assume it to be 
renormalized on-shell so that $\Pi_{\rm R}(0)=0$.  We consider
the expansion
\begin{equation}
\Pi_{\rm R}(q^2) = 
\Pi_{\rm R}^{(1)}(q^2) +
\Pi_{\rm R}^{(2)}(q^2) +
\Pi_{\rm R}^{(3)}(q^2) + \dots \,,
\end{equation}
where the superscript denotes the loop order.
Each order separately fulfills the subtracted dispersion
relation 
\begin{equation}
\label{disp}
\Pi_{\rm R}^{(n)}(q^2) =
\frac{q^2}{\pi}
\int_{4 m_e^2}^\infty \dd q'^2
\frac{{\rm Im} \, \big [ \Pi^{(n)}_{\rm R}(q'^2 + \ii \epsilon) \big ]}{q'^2 ( q'^2 - q^2 )} \,.
\end{equation}
where care has been taken to fulfill
the condition that  $\Pi_{\rm R}^{(n)}(q^2)$ needs
to vanish for zero $q^2$.
We scale the $n$th loop order as in Eq.~(11a) of Ref.~\cite{LaJe2024},
\begin{equation}
\label{Pscale}
\Pi^{(n)}_{\rm R}(q^2) =
\left( \frac{\alpha}{\pi} \right)^n 
P^{(n)}_{\rm R}(q^2) \,.
\end{equation}
When comparing to Eq.~(2.1) and Eq.~(3.9) of Ref.~\cite{On2022},
one realizes that the relation of our $n$th-loop 
function $\Pi^{(n)}_{\rm R}$ and the $\rho^{(n)}$
of Ref.~\cite{On2022} is
\begin{equation}
{\rm Im} \big [ \,\Pi^{(n)}_{\rm R}(q^2) \big ] = 
\left( \frac{\alpha}{4 \pi} \right)^n \rho^{(n)}(q^2) \,.
\end{equation}
which implies, in particular, that
\begin{equation}
{\rm Im} \big [ P^{(n)}_{\rm R}(q^2) \big ]  =
\frac{ \rho^{(n)}(q^2) }{2^{2n}} \,.
\end{equation}
The one-loop function ${\rm Im} \big [ P^{(n)}_{\rm R}(q^2) \big ]$ 
is given as 
\begin{equation}
{\rm Im} \big [ P^{(1)}_{\rm R}(q^2) \big ] =
\frac{\pi}{6} v \, (3-v^2) \, \Theta(q^2 - 4 m_e^2) \, .
\end{equation}
This result is well known and confirmed in 
Eq.~(28) of Ref.~\cite{LaJe2024} and 
Eq.~(3.1) of Ref.~\cite{On2022}, 
upon setting the number of fermion flavors 
$N=1$ in the latter and 
realizing that the $\beta$ variable, in
the notation of Ref.~\cite{On2022},
is equal to the Schwinger
$v$ parameter~\cite{Sc1970vol3}, which reads as
\begin{equation}
\label{defv}
v = \sqrt{ 1 - \frac{4 m_e^2}{q^2} } \,.
\end{equation}
The three-loop diagrams ($n = 3$) 
are depicted in Fig.~\ref{fig1}.
These are naturally divided into five categories.
The first two diagrams in the upper row 
of Fig.~\ref{fig1} are
vacuum-polarization insertions in the
inner virtual photon of the two-loop effect (first category).
The next category (the rightmost three diagrams in the
upper row) are two-loop self-energy insertions
in one of the fermion lines of the
one-loop effect.
The first two diagrams in the second row are self-energy
and vertex corrections to the one-loop vertex
correction diagram (fourth category).
The last three diagrams in the lower row are
generated from the one-loop vacuum-polarization
graph via the insertion of an equal number of
photon vertices in the upper and lower fermion lines (fifth category).

The imaginary part of the three-loop 
diagrams given in Fig.~\ref{fig1} 
can be written as
\begin{multline}
\label{ImP3R}
\Im \big [ P^{(3)}_{\rm R}(q^2 + \ii \epsilon) \big ]  = 
Q_{2m}(q^2) \, \Theta( q^2 - (2 m_e)^2 ) \\
+ Q_{4m}(q^2) \, \Theta( q^2 - (4 m_e)^2 ) \,.
\end{multline}
where the function $Q_{2m}(q^2)$ summarizes the terms 
with a two-fermion threshold, where the 
terms with a four-fermion threshold are summarized in $Q_{4m}(q^2)$. 
The former are obtained, 
for example, by cutting the third diagram in the
upper row of Fig.~\ref{fig1} right in the middle 
of the inner fermion lines, and using the 
Cutkosky rules~\cite{Cu1960}.
However, there are additional terms with a 
four-fermion threshold;
these are generated, for example, when one 
cuts the first Feynman diagram in the upper row
of Fig.~\ref{fig1} right in the middle.

The conversion to the conventions of
Ref.~\cite{On2022} is achieved by identifying
$Q_{2m}(q^2) = \frac{1}{64} \, \rho^{(3)}_{2m}(q^2)$
and $Q_{4m}(q^2) =  \frac{1}{64} \, \rho^{(3)}_{4m}(q^2)$.
For reference, we list the asymptotic behavior 
of these functions at their respective
thresholds $q^2 = (2 m_e)^2$ and $q^2 = (4 m_e)^2$,
and for large $q^2$.
The threshold expansion for $Q_{2m}(q^2)$ 
is most easily expressed in terms of the 
variable $v$, which is defined in Eq.~\eqref{defv},
\begin{multline}
\label{P2Mthr}
Q_{2m}(q^2) = \frac{\pi^5}{24 v} - \pi^3 -
\frac{\pi^3 v}{3} \, \ln( v )
+ v 
\biggl[ -\frac{\pi \, \zeta(3) }{2}
\\
+ \frac{5 \pi^5}{72}
- \frac{43 \pi^3}{36}
+ \frac{527 \pi}{72}
+ \frac{2 \pi^3}{3} \ln(2) 
\biggr] + \calO(v^2 \ln(v)) \,.
\end{multline}
where we take note of the fact that
$v=0$ at the threshold $q^2 = (2 m_e)^2$.
Alternatively, the first terms
read as follows, when expressed in terms
of $q^2/m_e^2 - 4$, which also goes to zero
at threshold, $Q_{2m}(q^2) = \frac{\pi^5}{12 \, \sqrt{ q^2/m_e^2 - 4}} 
- \pi^3 + \calO( \sqrt{ q^2/m_e^2 - 4} )$.
The asymptotics for high $q^2$ of 
$Q_{2m}(q^2)$ involve triple logarithms,
\begin{multline}
\label{P2Masy}
Q_{2m}(q^2) = 
- \frac{\pi}{54} 
\, \ln^3\left( \frac{q^2}{m_e^2} \right)
+ \frac{19 \pi}{108} 
\, \ln^2\left( \frac{q^2}{m_e^2} \right)
\\
+ \left( - \frac{881 \pi}{648} + \frac{13 \pi^3}{108} 
- \frac{\pi}{3} \zeta(3) \right) \,
\ln\left( \frac{q^2}{m_e^2} \right)
\\
- \frac{\pi}{3} \, \zeta(3)
+ \frac{19 \pi^5}{1080}
- \frac{\pi^3}{3} \ln (2) 
- \frac{89 \pi^3}{216}
+ \frac{15767 \pi}{2592}
\\
+ \calO\left( \frac{m_e^2}{q^2} \,
\ln\left( \frac{q^2}{m_e^2} \right) \right) \,.
\end{multline}
The four-fermion-threshold term 
$Q_{4m}(q^2)$ ramps up smoothly from its
threshold, without an (integrable) singularity,
\begin{multline}
\label{P4Mthr}
Q_{4m}(q^2) = 
\frac{11 \, \pi^2}{330301440} \, 
\left( \frac{q^2}{m_e^2} - 16 \right)^{9/2}
\\
- \frac{89 \, \pi^2}{11626610688} \, \left( \frac{q^2}{m_e^2}  - 16 \right)^{11/2}
\\
+ \frac{1055 \pi^2}{906875633664} \, \left( \frac{q^2}{m_e^2} - 16 \right)^{13/2}
\\
+ \calO\left( \left( \frac{q^2}{m_e^2} - 16 \right)^{15/2} \right) \,.
\end{multline}
The asymptotics 
of $Q_{4m}(q^2)$ for high $q^2$ cancel the 
double and triple logarithms from Eq.~\eqref{P2Masy},
\begin{multline}
\label{P4Masy}
Q_{4m}(q^2) =
\frac{\pi}{54}
\, \ln^3\left( \frac{q^2}{m_e^2} \right)
- \frac{19 \pi}{108}
\, \ln^2\left( \frac{q^2}{m_e^2} \right)
\\
+ \left( \frac{935 \pi}{648} - \frac{13 \pi^3}{108}
+  \frac{\pi}{3} \zeta(3) \right) \,
\ln\left( \frac{q^2}{m_e^2} \right)
\\
+ \frac{2 \pi}{3} \, \zeta(3)
- \frac{19 \pi^5}{1080}
+ \frac{\pi^3}{3} \ln (2) 
+ \frac{89 \pi^3}{216}
- \frac{4259 \pi}{648}
\\
+ \calO\left( \frac{m_e^2}{q^2} \,
\ln\left( \frac{q^2}{m_e^2} \right) \right)
\end{multline}
For large $q^2$, in view of a considerable cancelation
between $Q_{2m}(q^2)$ and $Q_{4m}(q^2)$,
there is only a single logarithm left,
\begin{multline}
\label{ImP}
{\rm \Im} \big [ P^{(3)}_{\rm R}(q^2) \big ] = Q_{2m}(q^2) + Q_{4m}(q^2) 
\\
= \frac{\pi}{12} \, \ln\left( \frac{q^2}{m_e^2} \right)
+ \frac{\pi}{3} \, \zeta(3)
- \frac{47 \pi}{96}
\\
+ \calO\left( \frac{m_e^2}{q^2} \,
\ln\left( \frac{q^2}{m_e^2} \right) \right) \,.
\end{multline}
From Eq.~\eqref{ImP}, with the help of the 
dispersion relation~\eqref{disp},
one may infer the leading logarithmic asymptotics
of $P^{(3)}_{\rm R}(-\vec q^{\,2}) $ for 
large spatial momentum transfer $q^2 = -\vec q^{\,2}$,
and compare with Ref.~\cite{FaKaLaSt1991}.
However, for the non-logarithmic term,
the calculation is more complicated.
The non-logarithmic term for the 
entire three-loop function is known, 
and can be inferred from equations 
presented in the text following Eq.~(4) of 
Ref.~\cite{Ka1992} 
(for the non-logarithmic term proportional to $N^2$) 
and from Eq.~(15) of Ref.~\cite{BrKaTa1992}
and Eq.~(4) of Ref.~\cite{BaBr1995}
(for the non-logarithmic term proportional to $N$).
(Note that the quantity $N$, the number of fermion
flavors, counts the number of electron loops in Fig.~\ref{fig1}.)
One may infer the result
\begin{multline}
\label{asymp}
P^{(3)}_{\rm R}(-\vec q^{\,2}) = 
-\frac{1}{24} \ln^2\left( \frac{\vec q^{\,2}}{m_e^2} \right) 
+ \left( \frac{47}{ 96 } - \frac{\zeta(3)}{3} \right) \, 
\ln\left( \frac{\vec q^{\,2}}{m_e^2} \right) 
\\
- \frac{1703}{1728} - \frac{23 \pi^2}{72} 
+ \frac{\pi^2}{3} \ln(2)
- \frac{173}{288} \, \zeta(3) 
+ \frac{5}{2} \, \zeta(5) 
\\
+ \calO\left( \frac{m_e}{\vec q^{\,2}} \,
\ln^2\left(\frac{\vec q^{\,2}}{m_e^2} \right) \right) \,.
\end{multline}
The numerical value of the non-logarithmic coefficient
is $0.012290603\dots$.
We have verified the asymptotic expansion~\eqref{asymp}.

Finally, the three-loop vacuum-polarization
correction to the Coulomb potential can be expressed
as a generalization of Eq.~(63) of Ref.~\cite{LaJe2024},
\begin{equation}
V^{(3)}_\rmR(r) = - \frac{Z \alpha}{\pi}
\int\limits_{4 m_e^2}^\infty \frac{\dd (q^2)}{q^2}
\; \frac{\ee^{- q r}}{r}
\; \mathrm{Im} \! \left[ \Pi^{(3)}_\rmR (q^2 + \ii \epsilon) \right] \,,
\end{equation}
where according to Eq.~\eqref{Pscale},
\begin{equation}
{\rm Im} \left[ \Pi^{(3)}_{\rm R}(q^2 + \ii \epsilon) \right] =
\left( \frac{\alpha}{\pi} \right)^3
{\rm Im} \big [ P^{(3)}_{\rm R}(q^2 + \ii \epsilon) \big ] \,,
\end{equation}
and ${\rm Im} \big [ P^{(3)}_{\rm R}(q^2) \big ]$ is given in 
Eq.~\eqref{ImP3R}.

%
%
\section{Numerical Calculation}
\label{sec3}

The energy shift
\begin{multline}
\label{E3}
E^{(3)}_{n \ell} =
\left< n \ell | V^{(3)}_\rmR(r) | n \ell \right> 
= - \frac{Z \alpha}{\pi}
\\
\times \int\limits_{4 m_e^2}^\infty \frac{\dd (q^2)}{q^2}
\; \left< n \ell m \left|  \frac{\ee^{- q r}}{r} \right| n \ell n \right>
\; \mathrm{Im} \! \left[ \Pi^{(3)}_\rmR (q^2 + \ii \epsilon) \right] \,,
\end{multline}
can be evaluated in first-order perturbation theory,
for a nonrelativistic state with principal quantum number
$n$ and orbital angular momentum quantum number $\ell$.
The energy shift is independent of the magnetic projection $m$. 
In writing Eq.~\eqref{E3}, we follow the conventions 
of Refs.~\cite{On2022,LaJe2024} for the sign of the 
vacuum-polarization function. When comparing to the 
(opposite) sign conventions used in Ref.~\cite{JeAd2022book},
one notices that in Eq.~(10.241) of Ref.~\cite{JeAd2022book},
the imaginary part of the vacuum-polarization
function is taken at $q^2 - \ii \epsilon$,
{\em i.e.}, below the cut, leading to consistency 
with Eq.~\eqref{E3}.
We have used the results of Ref.~\cite{On2022} for 
the three-loop  vacuum polarization spectral density 
$\mathrm{Im} \! \left[ \Pi^{(3)}_\rmR (q^2 + \ii \epsilon) \right] =\rho^{(3)}(q^2) $.
Specifically, for $q^2$ above and close to $(2 m_e)^2$, we 
have used the 
threshold expansion, which is a series in $v=\beta=\sqrt{1-\frac{4m_e^2}{q^2}}$,
and for large $q^2$ , we used the ``high--energy'' expansion, 
which is a series
in $m_e^2/q^2$.  These expansions are provided in supplementary files
accompanying Ref.~\cite{On2022}, containing exact results for the
first 120 terms in the threshold expansion and the first 50 terms in the
high-$q^2$ expansion.  For the point where we switched from use of
one series to the other, we chose the point where the two truncated series
best match each other, which is near $\bar q^2 = 8.56 \, m_e^2$.  At that point
they are both equal to $-235.325\, 929\, 481\, \dots$, with a difference
of a few parts in $10^{13}$.
 
It is convenient to define the (dimensionless) $\beta$ parameter,
\begin{equation}
\beta = \frac{m_e}{Z \alpha m_r} \,,
\end{equation}
which is equal to the product of the generalized Bohr
radius $a_0 = 1/(Z\alpha m_r)$ and the electron mass
(or, equivalently, in natural units, the 
inverse of the reduced electron Compton wavelength).
For muonic bound systems with nuclear 
charge numbers $1 \leq Z \leq 6$, the nuclear masses and $\beta$ 
parameters are (to six significant figures):
\begin{subequations} 
\begin{eqnarray}
m(\mbox{H}) &=& 938.272 \, {\rm MeV} \, , \; \;  
  \beta(\mu\mbox{H}) = 0.737384 \, ,  \\
m(\mbox{D}) &=& 1875.61 \, {\rm MeV} \, , \; \;  
  \beta(\mu\mbox{D}) = 0.700086 \, ,  \\
m({}^3 \mbox{He}) &=& 2808.39 \, {\rm MeV} \, , \; \;  
  \beta(\mu {}^3 \mbox{He}) = 0.343843 \, ,  \\
m({}^4 \mbox{He}) &=& 3727.38 \, {\rm MeV} \, , \; \; 
  \beta(\mu {}^4 \mbox{He}) = 0.340769 \, ,  \\
m({}^6 \mbox{Li}) &=& 5601.52 \, {\rm MeV} \, , \; \;  
  \beta(\mu {}^6 \mbox{Li}) = {\color{black} 0.225084} \, ,  \\
m({}^7 \mbox{Li}) &=& 6533.83 \, {\rm MeV} \, , \; \; 
  \beta(\mu {}^7 \mbox{Li}) = {\color{black} 0.224490} \, ,  \\
m({}^9 \mbox{Be}) &=& 8392.75 \, {\rm MeV} \, , \; \;  
  \beta(\mu {}^9 \mbox{Be}) = 0.167774 \, ,  \\
m({}^{10} \mbox{Be}) &=& 9325.50 \, {\rm MeV} \, , \; \;  
  \beta(\mu {}^{10} \mbox{Be}) = 0.167565 \, , \hspace{0.6cm} \\
m({}^{10} \mbox{B}) &=& 9324.44 \, {\rm MeV} \, , \; \;  
  \beta(\mu {}^{10} \mbox{B}) = 0.134052 \, , \\
m({}^{11} \mbox{B}) &=& 10252.5 \, {\rm MeV} \, , \; \; 
  \beta(\mu {}^{11} \mbox{B}) = 0.133916 \, ,  \\
m({}^{12} \mbox{C}) &=& 11174.9 \, {\rm MeV} \, , \; \; 
  \beta(\mu {}^{12} \mbox{C}) = 0.111503 \, , \\
m({}^{13} \mbox{C}) &=& 12109.5 \, {\rm MeV} \, , \; \; 
  \beta(\mu {}^{13} \mbox{C}) = 0.111422 \, . 
\end{eqnarray} 
\end{subequations}
We have calculated the nuclear masses using the formula
\begin{equation}
m \approx A \, {\rm u} - Z \, m_e + \Delta m \, ,
\end{equation}
where $A$ is the atomic mass number, $u$ the atomic mass unit, and values for
the ``mass excess'' $\Delta m$ were taken from the 
pertinent tables in Ref.~\cite{KoEtAl2021}.
These values are equal to the nuclear masses to the level
of precision required, since binding energies are negligible 
at the level of accuracy required for our studies.

In order to evaluate the energy shift~\eqref{E3},
it is advantageous to first calculate the matrix 
element of the operator $ Z\alpha \frac{\ee^{- q r}}{ r} $,
which can easily be done analytically.
We list results for states with maximum orbital angular
momentum $\ell$ for given principal quantum number $n$,
\begin{subequations}
\begin{align}
\langle 1S | Z\alpha \frac{\ee^{- q r}}{ r} | 1S \rangle =& \;
\frac{4 (Z\alpha)^2 m_r}{(2 + (q/m_e) \beta)^2 } \,,
\\
\langle 2P | Z\alpha \frac{\ee^{- q r}}{r} | 2P \rangle =& \;
\frac{ 4 (Z\alpha)^2 m_r}{(2 + (q/m_e) \beta)^4 } \,,
\\
\langle 3D| Z\alpha \frac{\ee^{- q r}}{r} | 3D \rangle =& \;
\frac{64 (Z\alpha)^2 m_r}{9 (2 + 3 (q/m_e) \beta)^6 } \,,
\\
\langle 4F | Z\alpha \frac{\ee^{- q r}}{r} | 4F \rangle =& \;
\frac{ 16 (Z\alpha)^2 m_r}{(2 + 3 (q/m_e) \beta)^8 } \,.
\end{align}
\end{subequations}
The results on the right-hand side involve a common
scaling factor $(Z\alpha)^2 m_r$ (the generalized
Hartree energy), and a residual dependence on the $\beta$ parameter,
which is of order unity 
for muonic bound systems of interest,
and a function of the dimensionless
ratio $q/m_e$, which, likewise, is of order
unity for the integration domain relevant to
one-loop, two-loop, and three-loop vacuum
polarization. Hence, one can understand why 
the three-loop vacuum-polarization effect is of 
order $(\alpha/\pi)^3 \, (Z\alpha)^2 m_r$
for muonic bound systems, and thus,
much less suppressed as compared to electronic 
bound systems.

Finally, the energy shifts $E^{(3)}_{n \ell}$
of Eq.~(\ref{E3}) due to irreducible three-loop vacuum
polarization are given in Table~\ref{table1}. 

The contribution of irreducible three-loop vacuum polarization
energy corrections have previously been calculated for several transitions in 
muonic systems.  For muonic hydrogen,
we can compare our results to 
Refs.~\cite{KiNo1999prl,KiNo1999prd}.
The reduced mass in this case is $m_r = m_\mu m_p/(m_\mu + m_p)$,
where $m_\mu$ is the muon mass, 
and $m_p$ is the proton mass.
From Eqs.~(18) and~(23) of Ref.~\cite{KiNo1999prl},
one infers the result
\begin{multline}
\left. E(2P) - E(2S) \right|_{\mu{\rm H}} 
= \Big [ 0.013628(6) \\
+ 0.017419(9) \Big ] 
\left. \left( \frac{\alpha}{\pi} \right)^3 \, 
(Z \alpha)^2 \, m_r \right|_{Z=1}
\\
= 0.0019671(7) \, {\rm meV} \,.
\end{multline}
We observe excellent agreement with the 
(numerically more precise) result
\begin{multline}
\left. E(2P) - E(2S) \right|_{\mbox{$\mu$H}} 
= \Big [ 2.6243(1) \times 10^{-3} \\
- 6.5673(1) \times 10^{-4} \Big ] \, {\rm meV} 
= 0.0019676(1) \,  {\rm meV} 
\end{multline}
from Table~\ref{table1}. The calculations in Refs.~\cite{KiNo1999prl,KiNo1999prd}
were done both from a direct numerical evaluation of the appropriate Feynman
diagrams and from use of a 
Pad\'e approximate developed in Ref.~\cite{BaBr1995}.
More recently, a calculation of the 3S-1S transition in 
muonic hydrogen gave a result of $0.0246 \,{\rm meV}$
(Ref.~\cite{DoEtAl2020}) for the irreducible 
three-loop vacuum polarization contribution, 
consistent with our more precise 
result $0.024390\,{\rm meV}$.  In addition,
results have been obtained for the 2S-1S 
transition in the muonic ions 
$\mu {}^7 {\rm Li}$, $\mu {}^9 Be$, 
and $\mu {}^{11} {\rm B}$ \cite{DoEtAl2021}.  Our results are consistent with 
all of these but more precise, in view of the 
avoidance of the numerical uncertainty inherent
to the Pad\'{e} approximants. The calculations 
reported in Ref.~\cite{DoEtAl2021}
made use of the Pad\'e approximants~\cite{BaBr1995}
for a function related to the irreducible three-loop 
vacuum polarization function for graphs
involving one electron loop only (all but the first 
two diagrams of Fig.~\ref{fig1}.
Six specific pieces
of information about the vacuum-polarization function 
along with its general analytical properties
were used to construct the approximation. The six items were 
the first three coefficients of the
expansion around $q^2=0$, the first two coefficients of 
the expansion in $1/q^2$ for large negative
$q^2$, and the the threshold behavior near 
$q^2=(2 m_e)^2$.  These six quantities were used to
construct Pad\'e $[3/2]$ and $[2/3]$ approximates, 
which were then used to obtain the vacuum-polarization function 
and the corresponding energy shifts.  The contributions of the first two
diagrams of Fig.~\ref{fig1}, having two electron loops, 
were computed separately.  
Our approach, making use of the first 120 terms in the 
threshold expansion and the first 50 terms in
the large negative $q^2$ expansion, gives an improved 
representation of the vacuum
polarization scalar function, leads to more precise results, 
and is straightforward to apply.

%
%
\section{Conclusions}
\label{sec4}

We have investigated the contribution of three-loop irreducible 
vacuum-polarization diagrams to the bound-state energy levels 
of muonic bound systems.  The evaluation of these corrections 
constitutes a step forward in the detailed knowledge
of the spectrum of these bound systems.
The three-loop corrections
are of order $\alpha^3 (Z\alpha)^2 m_r$.
It is instructive to compare their
numerical magnitude to another class of 
recently evaluated corrections, namely,
electronic vacuum-polarization
corrections to the self energy~\cite{WuJe2011,Bo2012,KaIvKa2013,%
KrMaMaFa2015,DoEtAl2019,DoEtAl2020,DoEtAl2021,PaEtAl2024,OhJe2024sevp}.
The latter are of order 
$\alpha^2 (Z\alpha)^4 m_r$,
and thus, formally, suppressed by an additional factor 
of $\alpha$ in comparison to the three-loop
vacuum-polarization corrections calculated here.
However, quite surprisingly, the corrections due to three-loop
vacuum polarization turn out to be 
of the same order-of-magnitude as the vacuum-polarization
corrections to the self-energy.
The observation becomes understandable
if one considers that the three-loop corrections
are suppressed by 
one power of $\pi$ more in the 
denominator, and a lack of 
a logarithmic enhancement factor $\ln(Z\alpha)$,
and a certain suppression in the numerical
coefficients multiplying the terms.
The three-loop corrections roughly scale only with 
the second power of the nuclear charge number,
$Z^2$, and are thus numerically most important
in comparison to other corrections for low 
nuclear charge numbers. 
Conversely, the three-loop corrections are 
numerically suppressed in comparison to other corrections
which scale with higher powers of the 
nuclear charge number $Z$.

The method of calculation used in this paper makes use of the irreducible
three-loop vacuum polarization spectral density, which has recently been
obtained analytically~\cite{On2022}, and specifically of its expansions
near threshold and for high negative $q^2$. It is relatively easy to
obtain the full vacuum polarization function through use of a subtracted
dispersion relation, and to verify its asymptotics given 
in Eq.~\eqref{asymp}. Our calculations remove the
theoretical uncertainty for the energy levels
of muonic bound systems with nuclear charge numbers
$1 \leq Z \leq 6$ due to irreducible
three-loop vacuum-polarization diagrams. These diagrams 
constitute, from a field-theoretical point of view, 
probably the most interesting and challenging subset of the 
$\alpha^5 \, m_r$ corrections.

\section*{Acknowledgments}

\vspace{-0.2cm}

This work was supported by the National Science Foundation through Grants
PHY-2308792 (G.S.A.) and PHY--2110294 (U.D.J.), and by the National Institute
of Standards and Technology Grant 60NANB23D230 (G.S.A.).

\end{document}